\documentclass[a4paper,twocolumn,11pt,accepted=2024-11-04]{quantumarticle}
\pdfoutput=1
\usepackage[utf8]{inputenc}
\usepackage[english]{babel}
\usepackage[T1]{fontenc}
\usepackage{amsmath}
\usepackage{hyperref}
\usepackage{physics}
\usepackage{tikz}
\usepackage{lipsum}
\usepackage{graphicx}                  
\usepackage{amssymb}
\usepackage{bm}
\usepackage{multicol}
\usepackage{array}
\usepackage{epstopdf}
\usepackage{mathrsfs}
\usepackage{grffile}
\usepackage{floatrow}
\usepackage[amssymb]{SIunits}
\usepackage{floatflt} 
\usepackage{wrapfig}
\usepackage{amsthm}
\usepackage{dsfont}
\usepackage{geometry}
\usepackage{setspace}
\usepackage{authblk}
\usepackage[numbers]{natbib}
\usepackage{dsfont}
\usepackage{breqn}

\newtheorem{assumption}{Assumption}

\newtheorem{definition}{Definition}

\begin{document}

\title{Epistemic Boundaries and Quantum Uncertainty: What Local Observers Can (Not) Predict}

\author{Johannes Fankhauser}
\thanks{}
\affiliation{Institute for Theoretical Physics, University of Innsbruck, Technikerstraße 21a, 6020 Innsbruck, Austria}
\affiliation{Faculty of Philosophy, University of Oxford, Woodstock Road, Oxford OX2 6GG, UK}
\orcid{0000-0001-8973-0373}
\email{johannes.fankhauser@uibk.ac.at}

\maketitle

\begin{abstract}
 One of quantum theory's salient features is its apparent indeterminism, i.e. measurement outcomes are typically probabilistic. We formally define and address whether this uncertainty is unavoidable or whether post-quantum theories can offer a predictive advantage while conforming to the Born rule on average. We present a no-go claim combining three aspects: predictive advantage, no-signalling, and reliable intersubjectivity between quantum observers. The results of the analysis lead to the conclusion that there exists a fundamental limitation on genuine predictive advantage. However, we uncover a fascinating possibility: When the assumption of reliable intersubjectivity between different observers is violated, subjective predictive advantage can, in principle, exist. This, in turn, entails an epistemic boundary between different observers of the same theory. The findings reconcile us to quantum uncertainty as an aspect of limits on Nature's predictability.
\end{abstract}

\section{Introduction}

Insofar as the formalism is deemed valid, quantum theory's predictions are generally probabilistic. That is, generic physical states imply uncertainty for the outcomes of measurements. At face value, the quantum formalism suggests that its probabilism is unsurmountable, so quantum indeterminism has become one of the characteristics of the theory. 

Quantum theory's indeterminism can be articulated in terms of a `decorrelation principle' \cite{Fankhauser-thesis}: This principle states that when a pure state is assigned to a target system $S$, the outcomes of any measurement are statistically independent of its environment. Consequently, the maximum information that the environment can contain is limited to the information about the target system's pure quantum state:
\begin{equation}
	\label{eqn:decorrelation_principle}
	p(x,y|\rho)= p(x|\rho_S)\cdot p(y|\rho_E),\nonumber
\end{equation} where $x,y$ are the outcomes of the measurements on system $S$ and the environment $E$, $\rho$ the total joint state, and $\rho_S, \rho_E$ the respective reduced states.\footnote{Classical mechanics can be said to obey a similar decorrelation principle. However, the `pure' states are the exact positions and momenta of particles which together with the physical laws contain no uncertainty about the future state of a system. For a caveat to this common claim see \cite{epistemic_horizon}.}

The question arises whether we expect quantum uncertainty to be fundamental in future theories as well, or whether there could exist non-quantum variables with predictive advantage.

But why does the quantum formalism suggest that there is a limit to the predictability of an event? Is it because Nature is \textit{fundamentally} indeterministic?  Still, if Nature were deterministic, this raises the question of why it shouldn't be \textit{predictable}. Can quantum uncertainty be a consequence of a more fundamental physical principle? 

Apart from the foundational significance, the question has practical relevance and consequences for post-quantum theories: What is the most accurate prediction possible for physical phenomena? What are physical principles independent of the quantum formalism that can be used to make statements about the predictability in post-quantum theories. 

In 1935 the puzzle about quantum theory's completeness culminated in the notorious `EPR-argument': Imagine a widely separated pair of entangled particles that can thus no longer interact with one another. In turn, the situation is said to satisfy Einstein, Podolsky, and Rosen's element of reality: Since ``without in any way disturbing the system''  (as the particles are space-like separated) ``we can predict with certainty (i.e. with probability equal to unity) the value of a physical quantity, [...]  there exists an element of physical reality corresponding to this physical quantity.'' \cite[p.~777]{einstein1935can}. Since the particles are entangled, one can therefore simultaneously perform a position measurement on one and a momentum measurement on the other particle. If the remote measurement outcome can be predicted with certainty (due to the perfect quantum correlations) without interfering with the system (by assuming the validity of the theory of relativity), the outcome must be pre-determined and thus the quantum description is incomplete: Einstein was ``[...] firmly convinced that the essentially statistical character of contemporary quantum theory is solely to be ascribed to the fact that this [theory] operates with an incomplete description of physical systems''. \cite[p.~666]{einstein1935can}

Einstein played with other thought experiments to conceive of scenarios violating Heisenberg's uncertainty relations (see e.g. perpetua mobilia like the Einstein paradox, \cite{Bohr-Schilpp1949}). He knew that knowledge of the uncertainty principles doesn't imply that a more complete theory is impossible. Einstein was primarily concerned with the ontological rather than predictive completeness of quantum mechanics. EPR never claimed that the theory would be predictively complete (see, for instance, \cite{Schilpp1949}).

Important recent results by Barrett, Kent, and Pironio \cite{Barrt-Kent-Pironio-local-decomposition} as well as Colbeck and Renner \cite{colbeck-renner} addressed predictive completeness and showed that quantum predictions cannot be extended by employing a version of chained Bell inequalities. 

However, these results do not take into account the significance of reliable intersubjectivity, i.e. the shared experience of measurement records by macroscopic observers. Here we demonstrate that reliable intersubjectivity is a crucial and necessary assumption. It can be seen as a generalisation of what is commonly known as `absoluteness of observed events’ which plays a crucial role in recent results involving multi-agent scenarios.\cite{Cavalcanti-local-friendliness-no-go, Brukner-Wiger-no-go, Frauchiger-Renner-no-go, Brukner-persistent-reality-of-Wigner, Healey-limits-of-objectivity} None of these (extended) Wigner's friend type experiments~\cite{Wigner-measurement-problem}, however, addresses the question of predictive advantage.

We present a novel argument for the predictive completeness of quantum theory based on no-signalling and reliable intersubjectivity. This is achieved by giving a definition of predictive advantage through a violation of empirical completeness. The definition is purely operational and avoids the introduction of an ontological model \cite{Harrigan-Spekkens-2010}. We prove that the quantum state provides complete information about the predictions of measurement outcomes. More concretely, no predictive advantage can \textit{objectively} exist for all observers of a theory that cannot signal superluminally while reproducing the quantum probabilities on average. However, predictive advantage is indeed possible in the presence of \textit{epistemic boundaries}: The potential predictive advantage of one observer cannot be shared with other observers. 

To our knowledge, this is the first rigorous result on predictive completeness addressing locality and reliable intersubjectivity.

We proceed as follows. Section \ref{sec:empirical_completeness} introduces the relevant notion of empirical completeness and the setup of the main result. In Section \ref{sec:results} we formalise the assumptions of the theorem. The findings are discussed in Section \ref{sec:discussion}, where we comment on related results and the implications for interpretations of quantum theory. The proof of the theorem is given in Appendix \ref{app:methods}. We conclude in Section \ref{sec:conclusion}.
	
\section{Predictions of No-Signalling Observers}
\label{sec:empirical_completeness}

Let's begin with an instructive example from classical physics. Imagine a classical thermodynamic system of a gas in a box to illustrate how empirical \textit{in}completeness may turn out. The box of gas may be described as a closed system with a constant temperature. Subject to kinetic gas theory, a measurement of the velocity $v$ of the gas particles in the ensemble is expected to give values distributed according to Maxwell-Boltzmann statistics $f_{MB}(v)$, peaked around some mean velocity. Thus, the velocity $v$ of an individual particle will be found with some likelihood $p(v)$. But let’s imagine the outcomes of many repeated measurements on the same gas yield doubly peaked distributions corresponding to two Maxwell-Boltzmann distributions of different mean velocities. Let’s assume the two peaks originate from the fact that there are two distinct gases in the box, namely nitrogen and oxygen. Since the two gases have different kinetic properties, this leads to separate mean velocities. But to an experimenter without a physical theory distinguishing the two molecules, only \textit{one} gas is in the box. Thus, the experimenter’s original theory is empirically incomplete since the measurement outcomes will show an in-principle uncertainty resulting from treating the gas as one single substance. A more advanced theory that can distinguish nitrogen from oxygen will obtain more complete predictions.\footnote{For this to be possible, an actual experiment, i.e. preparation, has to exist that distinguishes nitrogen from oxygen, such that, on average, the total distribution  $f_{MB}(v)$ is still recovered. This is crucial, for otherwise, the improved theory wouldn't be an empirical completion but rather a metaphysical completion.}


The example shows how a theory, making probabilistic predictions, can be replaced by an improved theory with refined predictions. One may ponder if a similar description is conceivable for quantum theory. We define empirical completeness of a theory as follows (cf. also \cite{Fankhauser-thesis}): 
\begin{definition}{\textbf{Empirical Completeness.}}
	\label{def:empirical extension}	
	If for two theories $T_1$ and $T_2$ the \textit{same} system is described by states represented by configurations $y_1\in \mathcal{M}$ being from $T_1$, $y_2\in \mathcal{M}$ being from $T_2$, and it holds that $p(x|y_1)\neq p(x|y_2)$ for some outcome $x$, then $T_2$ is said to be an empirical extension or empirical decomposition of $T_1$ if
	\begin{equation}
		p(x|y_1)=\int p(x|y_2)\rho(y_2|y_1)d y_2,
	\end{equation} where $\rho(y_2|y_1)$ is some probability distribution over the target system states of theory $T_2$ for fixed preparation $y_1$. When there exists no \textit{empirical} extension for a theory $T_1$, it is called \textit{empirically complete}.
\end{definition} 
All variables involved---outcomes $x$, and preparations $y_1,y_2$---are configurations in the manifest domain $\mathcal{M}$. The set $\mathcal{M}$ is a specification of the directly accessible empirical `data' on which an observer's knowledge supervenes\footnote{Using the standard notion of supervenience, this means that there cannot be a difference in the observers' knowledge state without a difference in the empirical data.} (cf. \cite{Fankhauser-thesis} for details).\footnote{It is worth noting that indeed \textit{every} probabilistic physical theory can be supplemented by a deterministic underpinning.  (for any sequence of events there exists a description that contains all information in the initial state of the system, cf. e.g. \cite{Fankhauser-thesis}, \cite[Chapter~3.4.1]{bricmont2016making}. A viable version of determinism will thus involve restrictions on the domain of variables allowed and the complexity of laws invoked). For instance, quantum theory is explicitly probabilistic, but pilot wave theory provides a deterministic underpinning.}


The idea to connect (versions of) locality with quantum completeness dates back to Einstein, Podolsky and Rosen, who argued for the incompleteness of quantum theory based on the existence of entangled states and their non-local properties \cite{einstein1935can}. If quantum measurement outcomes aren't further pre-determined by more than the probabilistic quantum prediction, they must be non-local (in the sense of EPR\footnote{Let $\rho$ be the state of a space-like separated bi-partite quantum system with measurement results $x$ and measurement settings $a$ on one side, and outcomes $y$ and settings $b$ on the other. Let $\lambda\in\Lambda$ be the supplementing hidden states of the system. Following Shimony \cite{Shimony1986,Shimony1990}, two notions of locality are distinguished. \textit{Parameter Independence:} \begin{align}
		p(x|a,b,y,\rho,\lambda)&=p(x|a,y,\rho,\lambda), \\ \nonumber
		p(y|a,b,x,\rho,\lambda)&=p(y|b,x,\rho,\lambda), \nonumber
	\end{align} for all $x,y, a,b, \lambda$. That is, the outcome probabilities are independent of the settings of the remote measurement. \textit{Outcome Independence:} \begin{align}
		p(x|a,b,y,\rho,\lambda)&=p(x|a,b,\rho,\lambda), \\ \nonumber
		p(y|a,b,x,\rho,\lambda)&=p(x|a,b,\rho,\lambda), \nonumber
	\end{align} for all $x,y, a,b, \lambda$, i.e. the outcome probabilities are independent of the outcomes of the remote measurement.
	If there exists no hidden state $\lambda$, then outcome independence effectively coincides with EPR's notion of locality. The conjunction of both parameter and outcome independence is needed to derive Bell's theorem.}) to recover the perfect correlations of entangled states \cite{maudlin2014bell}. Supplementing the quantum description with so-called hidden variables could then perhaps restore determinism, thus completing the theory to avoid what is often called `action at a distance'.

But by Bell's theorem, any such completions must be non-local too.\cite{bell-theorem} To avoid the paradoxical challenges of superluminal signalling this implies that if hidden variables exist, the standard theory can either not be completed, or some mechanism must exist limiting the extent to which such variables are empirically accessible. Otherwise, if experimenters could control hidden variables, they could employ them to signal. A generalisation of this claim is used in the main result of this paper.


The basic setup of the thought experiment used in the main result consists of three observers Alice, Bob, and Maggie which we shall abbreviate with $A, B$, and $M$. The source $S$ emits a pair of spin-$\frac{1}{2}$ particles prepared in the pure entangled state $\rho=\ket{\psi}\bra{\psi}$ with $\ket{\psi}:=\alpha\ket{00'}+\sqrt{1-\alpha^2}\ket{11'}$, where $\{\ket{00'}, \ket{11'}\}$ the Schmidt basis of the bi-partite system, and $\alpha\in \mathbb{R}$ a parameter to quantify the degree of entanglement. Alice and Bob perform measurements in the settings $a, b$ and obtain binary outcomes $x, y\in \{0,1\}$. 

\begin{figure*}[t]
	\centering
	\includegraphics[width=0.6\linewidth]{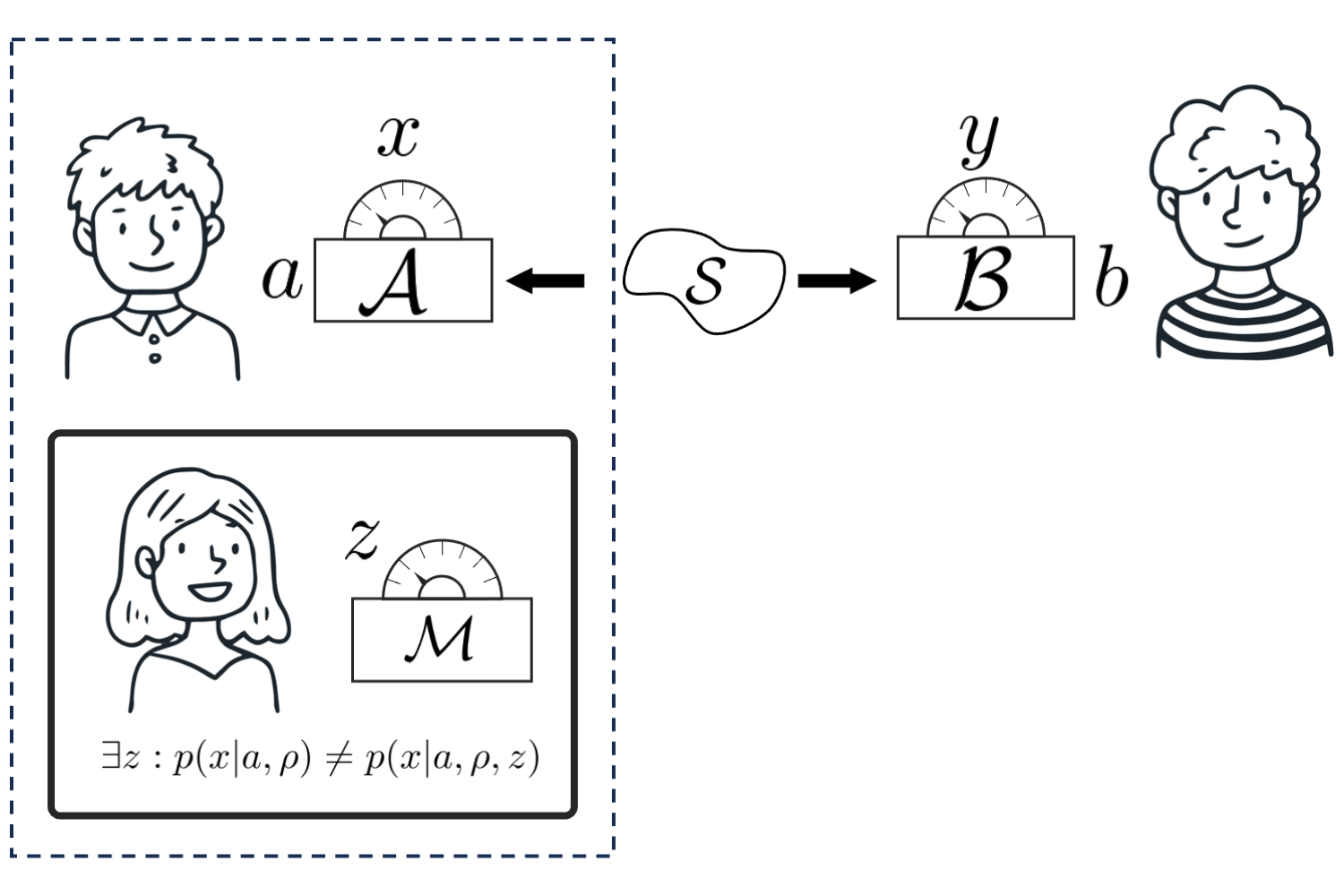}
	\caption{The experimental setup consists of three observers Alice, Bob, and Maggie ($A, B$, and $M$). The source $S$ emits a pair of entangled quantum particles. Alice and Bob perform measurements in the settings $a, b$ and obtain binary outcomes $x, y\in \{0,1\}$. Maggie obtains a variable $z$ with which she can make better predictions about Alice's outcomes than the quantum Born rule.}
	\label{fig:predictive-advantage-setup}
\end{figure*}

The third observer possesses a magic prediction box with outcome variables $z$ conditioned on which the predictions of Alice are hypothetically improved. That is, in each run of the experiment the probabilities for outcomes $x$ given the state $\rho$ may differ from the probabilities for given the state $\rho$ plus the value of $z$. The experiment is depicted in Figure \ref{fig:predictive-advantage-setup}. In Section \ref{sec:results} a precise assumption will be made to what extent $M$ can share with $A$ the value of $z$. 

In the next section, the assumptions needed for the theorem are discussed, and the main result is presented. It is shown under what conditions quantum mechanics can be considered predictively complete. No-signalling is shown to suffice to infer empirical completeness for maximally entangled quantum states. However, introducing epistemic boundaries between different users of the same theory---viz. violating the assumption of `reliable intersubjectivity', one can in principle get around limitations on predictive advantage.

\section{Results}
\label{sec:results}

Here we formulate the main claim and the assumptions on which it rests. We shall first introduce the auxiliary assumption on the nature of the variables involved to make predictions, i.e. define the variables which are thought of as being `accessible' to observers on which, in turn, predictions supervene. We then make precise what it means for an observer to have predictive advantage, and explicate the necessary assumption on the epistemic relationship between different observers of the theory, which we dub `reliable intersubjectivity'. This then allows us to make sense of the crucial assumption of no-signalling and the additional constraint of `no-conspiracy' which then serve to derive the main claim of this work. That is, the assumptions cannot all be jointly true. 

\subsection{Formalisation of the Assumptions}

The physical world can be taken to decompose into two domains. One contains the empirical data, i.e. the manifest elements like outcomes of measurements and records in the environment. It comprises what is directly observable as outputs, settings and physical data---the bare phenomenological facts.\footnote{For instance, these could be (but need not be) framed in terms of positions of objects, events, or configurations in 3 or 4-dimensional space. For an account of the manifest domain in terms of observable facts ultimately encoded in spatial degrees of freedom, i.e. as manifest configurations like ink on paper, points on a screen, or the position of a pointer see, for instance, \citep{Fankhauser-thesis}.} 

The manifest domain alone doesn't entail any \textit{a priori} commitment to what is real in the physical world beyond itself, i.e. what the elements in the manifest domain may or may not signify. For instance, a configuration of dots on a screen does not---without further theorising---imply the existence of a particle in this very spatial location. Theories about what is contained in the non-manifest domain will be informed by what is observed as the manifest domain.

The non-manifest domain contains the elements, objects and facts that \textit{aren't} directly observable. These will be the things that exist to which a physical theory relates the manifest elements. For example, it will tell when a manifest element, e.g. a position measurement record, actually refers to a non-manifest element, e.g. the position of a particle. We may label those elements with the familiar attribute of being `ontological'. 

This distinction is key. Discussions on the meaning of modern physics theories have demonstrated the relevance of making the manifest/non-manifest distinction. For example, much of the confusion in quantum paradoxes is due to misunderstandings about what measurement outcomes signify in quantum mechanics (see, for instance, \cite{Fankhauser-thesis}). Moreover, various modern results on the hidden variable programme showed that what may hold on the macroscopic, operational level doesn't necessarily carry over to the microscopic level.\footnote{For example, standard accounts of hidden variables theories do not give rise to superluminal signals on the \textit{manifest} level, although Bell's theorem implies that such theories  are necessarily non-local on the \textit{non-manifest} level.} 



In the context of our setup, there is thus the question of what the variables are that Alice, Bob, and Maggie can operationally access, manipulate, and empirically use to condition their physical predictions. By definition, we suppose that the set of variables accessible to observers is contained in the manifest domain. This entails the first assumption necessary to formulate the main claim:

\begin{assumption}[\textbf{Supervenience Domain (SD)}]
	\label{assumption:supervenience}
	There exists a set of (operationally accessible) manifest variables on which all predictions supervene for all observers. 
\end{assumption}

More concretely, applied to the present scenario, the manifest configurations are the settings and outcomes of the three agents, i.e. $a, b$, and $x, y, z$. 

Being clear on the status of the variables involved explicates that the predictively advantageous configurations $z$ available to Maggie are not `hidden variables' in the usual sense. They are assumed to be of the same nature as all other variables. Nevertheless, this is not to say that a theory may pose further variables that exist on top of what is accessible by the observers involved. 

This allows us to make more precise our second assumption. Consider the setup introduced above (Figure \ref{fig:predictive-advantage-setup}) with a bi-partite system with two observers Alice ($A$) and Bob ($B$) and outcomes  $x, y$ and measurement settings $a, b$. The prepared quantum state is assumed to be a fixed pure state $\rho$.

The third observer Maggie ($M$) possesses a magic prediction box providing predictive advantage over the quantum predictions for the entangled state. 

\begin{assumption}[\textbf{Predictive Advantage (PA)}]
	\label{assumption:predictive advantage}
	There exists a non-trivial decomposition of the quantum probabilities for at least one preparation and measurement, i.e.
	\begin{align}
		p(x,y|a,b,\rho) &=\tr(E_x^a\otimes E_y^b \rho) \\ \nonumber
		&= \int p(x,y|a,b,\rho,z)\mu(z)dz.
	\end{align}
\end{assumption} The quantum state $\rho$ leads to a preparation $\mu(z)$ for Maggie's prediction box. In principle, $z$ can also depend on the settings of the other observers which would lead to a version of superdeterminism, and retro-causation. The no-conspiracy Assumption \ref{assumption:no-conspiracy} below rules out such possibilities. $(PA)$ formalises the violation of empirical completeness \ref{def:empirical extension}.

The predictive advantage is non-trivial if for at least one value of the variable $z$ coming from the magic prediction box, the prediction deviates from the Born rule for at least some outcome $x$, setting $a$, and state $\rho$, i.e. (omitting $z$'s dependency on other variables)
\begin{dmath}
	\exists z: p(x|a, b,\rho)=\sum_y p(x,y|a,b,\rho) \neq  \sum_y p(x,y|a,b,\rho,z)=p(x|a,b,\rho, z).
\end{dmath}

Maggie is not assumed to perform standard quantum measurements on the shared entangled pair of particles. Thus, the setup of the main claim is different from familiar (extended) Wigner's friend experiments where it is standardly assumed that some observers (the friends) measure the quantum system and evolve quantum mechanically before super-observers do further quantum operations on their labs.\cite{Brukner-Wiger-no-go, Cavalcanti-local-friendliness-no-go, Frauchiger-Renner-no-go} Maggie is hence also not assumed to remain in a superposition of states that is passed on to Alice. We treat the problem as a question about what (informational) relevance the variable $z$ can have for Alice's outcomes without any assumption what the quantum state, if any, Maggie is in.

Predictive advantage is the assumption that relates her variable to the other variables in the setup. Either Maggie makes the same predictions as Alice---the standard Born probabilities---or, she makes improved predictions in which case $z$ acts as a source for pre-determination of Alice's outcomes.\footnote{Relatedly, in a recent review of Wigner's friend type arguments Leifer et al. point out that the results show a contradiction between the Born rule (i.e. the frequency by which outcomes will be observed without assumptions about the post-measurement state) and unitarity (and not just a tension between unitary evolution and possible state-update rules like the projection postulate)---except for the Frauchiger-Renner argument where the contradiction arises from combining Heisenberg cuts at different levels of observation (see, for example, \cite{Lazarovici-Huber-on-FR, Leifer-EWF-review}).}

The definition shifts the focus away from the ontological underpinning of the theory---such as e.g. in hidden variable theories---to prediction-making.\footnote{Note that there exists a similar but distinct notion of completeness which is concerned with the question of whether the quantum state is ontic. In other words, whether the same state of reality is compatible with more than one wavefunction. (see Pusey, Barrett, and Rudolph's seminal result on the reality of the quantum state. \citep{PBR-theorem})}. We are thus not concerned with the \textit{ontological} or \textit{metaphysical} status of the quantum state, but rather with its epistemic role as a means for predicting outcome probabilities.

This supports the need for a clear distinction between what's manifest, i.e. empirically accessible, from what isn't, i.e. non-manifest. The variables that may determine a single measurement result don't necessarily coincide with the ones utilisable for predicting that result. Rather than to attempt restoring determinism in quantum mechanics, the question is whether quantum mechanics is \textit{empirically} complete.

So far no assumption has been made as to whether $z$ is a well-defined variable for \textit{all} observers of the setup. Although Maggie might have predictive advantage over Alice conditioned on $z$, it is unclear if its value can be reliably communicated to Alice. 

Arguably, an objective predictive advantage supposedly exists only if Maggie's predictions are absolute in the sense that all observers agree on a single value for $z$ which can be reliably attained by at least Maggie and Alice. Moreover, if Maggie wasn't able to report $z$ to other observers, she would be epistemically separated from her environment, and thus she couldn't test her predictions empirically.  This leads to the next assumption. 

\begin{assumption}[\textbf{Reliable Intersubjectivity (RI)}]
	\label{assumption:reliable intersubjectivity}
	Observed measurement records have well-defined values for all observers and can be reliably reported. 
\end{assumption}

In the context of our thought experiment $(RI)$ is the assumption that the variables of the setup exist objectively in the sense that they are single well-defined values for all observers and the events of the manifest domain are accessible to all observers in an absolute and reliable fashion. 

Moreover, the assumption entails that Maggie's records can be trusted. When Alice asks Maggie about the value of $z$, it is reliably reported by her: What she observed at the time when the magic prediction box indicated a value for $z$ is also the value that Alice will observe when communicated by Maggie at a later time. 

$(RI)$ is what gives the variable $z$ its relevant meaning and is needed to define the objective probability distribution and its marginals, i.e.
\begin{equation}
	p(x|a, b, \rho, z) = \sum_y p(x,y|a,b,\rho,z).
\end{equation}
The need for reliable intersubjectivity can be framed as reminiscent of the quantum measurement problem.

Suppose that Maggie did in fact have some predictive advantage over Alice by virtue of her variable $z$, as a result, introducing a stronger pre-determination of Alice's measurement outcome. But this in turn could be taken as signifying the existence of an additional variable---represented by $z$---playing the role of a `hidden' variable (at least in the sense relative to Maggie). According to Maggie the quantum state is thus an incomplete description of the physical system, for the magic prediction box provides a more fine-grained description.

Importantly, reliable intersubjectivity does not introduce a hidden variable assumption though. On one hand, relative to Maggie and Alice the variable $z$ can be viewed as an additional variable completing the quantum state description. But note that $z$ is an observed record and hence not hidden.

Similar assumptions often appear in the context of (variants of) Wigner's friend experiments---the macroscopic version of the quantum measurement problem. \cite{Wigner-measurement-problem} There a macroscopic observer (the friend) exists in a superposition of states according to an external observer (Wigner). In terms of prediction making Wigner's original thought experiment highlights the tension between Wigner's predictions about the (superposed) friend's outcome, and what the friend may predict (upon having observed a definite outcome already). 

One natural proposal to (partly) accommodate the apparent ambiguity is to assume one particular choice of Wigner's measurement setting (the one with which the original thought experiment is standardly concerned). When he measures in the friend's basis, i.e. simply asks the friend to report her outcome,  then the measurement results are assumed to be compatible. In other words, Wigner essentially `checks' the friend's reading, and it seems plausible that Wigner's result equals what the friend observed. Occasionally this assumption is called `shared facts' (e.g. in Adlam's approach \citep{Adlam-intersubjectivity}). Moreover, this also follows from the so-called `absoluteness of observed events' $(AOE)$ and `observer-independent facts' (as employed in the extended Wigner's friend experiments by Brukner et al. \citep{Brukner-persistent-reality-of-Wigner} and Cavalcanti et al. \cite{Cavalcanti-local-friendliness-no-go}).\footnote{Note that intersubjectivity is distinct from the `repeatability' of quantum measurements. If the same measurement is performed twice, the results are in agreement. It also implies that if Wigner performed the same measurement as the friend \textit{directly} on the target system, their values would come out identical.  Compare, for instance, with what Maudlin calls `the problem of effect' in the context of the quantum measurement problem.\citep{maudlin1995three} The invoked consistency is a feature of the empirical content of the traditional theory and is experimentally well-confirmed.}

But $(RI)$ is more general than these assumptions. First, we don't assume Maggie is a superposed observer or performs a certain measurement on the entangled particles. It is only assumed that when conditioning on the definite variable $z$ she has \textit{some} predictive advantage over the predictions of Alice.

Applied to our scenario Brukner and Cavalcanti et al.'s assumption amounts to the special case where there exist deterministic probability assignments of Maggie's predictions for one choice of Alice's measurement, i.e.  $p(x,y|a,b)=\sum_z\delta_{x,z}p(x,y|b)p(z)$, when $a=1$---the measurement basis where Alice checks Maggie's outcome.\footnote{In this case it also bears resemblance to variants of what was coined in the literature as `macroscopic realism'---`the idea that a macroscopic system is always determinately in one or other of the macroscopically distinguishable states available to it.' \cite{Maroney-Timpson-on-macrorealism} It is worth mentioning that reliably intersubjectivity is prerequisite to EPR's criterion of an element of reality. For the values for $z$ are assumed not to be disturbed when reported, and Maggie can predict with certainty that Alice's observed value matches the value for $z$. One could thus argue that the assumption entails a common objective `reality' accessible by all observers.} 


Adlam and Rovelli follow a similar line of reasoning. In their recent work, the authors define a notion of `cross-perspective links' which is supposed to establish a consistency of measurement outcomes among different observers in relational quantum mechanics. \citep{Adlam-Rovelly-information-is-physical}). That is, when some observer obtains the value of a physical variable producing a record of that value, and another observer subsequently measures that same record, the two results should match.

This goes in the direction of reliable intersubjectivity as defined in this work. Note, however, that Adlam and Rovelli's definition is ambiguous in terms of whether the second measurement result is a \textit{reliable} record of the first measurement result. The two values can match after the second interaction, but this doesn't imply that at the time of the first measurement, the outcome had the same value. Moreover, the claim established in this paper might provide a counterargument to `cross-perspective links'. For their theory seems to accept all the other assumptions of our result which in turn implies that reliable intersubjectivity cannot hold in relational quantum mechanics. 

The assumptions mentioned apply for a single choice of measurement only (i.e. when Wigner checks the friend's reading) and therefore prove insufficient to account for the more general assumption $(RI)$. Alice performs measurements in an arbitrary basis, for which the same question arises: What can Maggie predict about Alice's outcomes conditioned on $z$? 

$(RI)$ and $(PA)$ stay away from any commitment to whether Maggie's outcome $z$ has anything to do with the `actual' state of the physical system when it's a superposed state according to Alice. Upon $(RI)$ the variable $z$ can be objectively communicated among different observers, but it is only by virtue of $(PA)$ that $z$ assumes the meaning of a variable `completing' the quantum state. Furthermore, we suppose $z$ to signify an \textit{arbitrary} predictive advantage. That is, $z$ is not constrained to deterministic predictions for one particular setting (e.g. the `record checking' measurement), or in fact to any deterministic prediction at all.

\begin{assumption}[\textbf{No-Signalling (SL)}]
\label{assumption:no-signalling}
The outcomes of an observer $A$ are conditionally independent of the settings of another space-like separated observer $B$ even when conditioned on other outcomes not in the future light cone of $A$.  
\end{assumption}

Upon reliable intersubjectivity, the record of variable $z$ can be observed by both Alice and Maggie. Hence, both observers will objectively agree on the single value of $z$, and the process of Alice attaining this record does not affect $z$. As a result, since Alice is in a position to access and employ $z$, the no-signalling condition has to hold when conditioned on the variable $z$. Otherwise, Alice could choose her settings depending on $z$ to send superluminal signals to Bob.  

Mathematically, this reads
\begin{align}
p(x|a,b,z)&= p(x|a,z) \\ \nonumber
p(y|a,b,z)&= p(y|b,z), \nonumber
\end{align} and this is non-trivial in the case where $z$ provides predictive advantage via Assumption \ref{assumption:predictive advantage}. No-signalling is a weaker notion of locality than Bell's parameter independence for hidden variables since if one is to propose the existence of such variables they may still be parameter dependent without $z$ leading to violation of no-signalling. Secondly, the assumption applies to actually observed events of Maggie. All the variables in the statement of the assumption are manifest configurations in the supervenience domain. 
No-signalling provides the last assumption necessary for the main result, together with one auxiliary condition, namely 

\begin{assumption}[\textbf{No-Conspiracy (NC)}]
\label{assumption:no-conspiracy}
Measurement settings are independent of outcomes outside their future light cone. 
\end{assumption}

Alice and Bob should be able to choose their measurements freely and independently of the quantum system (which includes the specification of Maggie's variable $z$). It can be assumed that the settings $a$ and $b$ are outside of the future light cone of $z$. That is, we have $p(a,b|z)=p(a,b)$, and by the definition of conditional probabilities the assumption $(NC)$ reads 
\begin{equation}
p(z|a,b) = p(z).
\end{equation} In the context of hidden variables theories $(NC)$ is also known as the `freedom of choice', `no-superdeterminism' condition, or setting independence.\footnote{For further considerations on this assumption see \cite{sep-bell-theorem}.}

We are now ready to state the main result of this work, which is the claim that assumptions (1)--(5) cannot all jointly be true. That is,
\begin{equation}
\label{eqn:main result}
\bm{(SD)  \wedge (PA) \wedge (RI) \wedge (SL) \wedge (NC) \rightarrow \bot.}
\end{equation}
The proof of this claim is found in Appendix \ref{app:methods}.

\section{Discussion}
\label{sec:discussion}
Drawing on quantum non-locality by generalised Bell inequalities for maximally entangled states, we have shown that whenever there would exist \textit{any} predictive advantage, observers could superluminally signal unless reliable intersubjectivity (or no-conspiracy) is violated. 

Here we shall discuss related questions and results. We first draw conclusions for the `absoluteness of observed events'. We then comment on the Colbeck-Renner theorem and then discuss the possibility of violating reliable intersubjectivity. We conclude with a brief discussion of pilot wave theory, Everettian quantum mechanics, and a note on Laplace's demon. 


\paragraph{Absoluteness of Observed Events.} 

This work advances existing related results involving Wigner's friend-type scenarios. As mentioned, the assumption of $(AOE)$ in the no-go theorem by Cavalcanti et al.~\cite{Cavalcanti-local-friendliness-no-go} can be reframed as a special case of reliable intersubjectivity and predictive advantage. Their claim follows from restricting the definition of predictive advantage to the case where Maggie's probabilities conditioned on $z$ yield a deterministic prediction for one of three measurement settings employed in their setup. That is, $(AOE)$ amounts to reliable intersubjectivity together with a version of predictive advantage. In the context of this work, Bong et al.'s no-go theorem shows that a certain measurement scenario can be found ruling out deterministic predictive advantage for one of three measurement settings  (barring the possibility of violating any of the other assumptions in the theorem). 

Assumption \ref{assumption:predictive advantage} is more general than $(AOE)$ in the sense that it allows for predictive advantage of \textit{any} kind for arbitrary settings and without any commitment to deterministic or probabilistic predictions for Alice's outcomes. This is made possible by employing the full non-locality of a maximally entangled state by using chained Bell inequalities, whereas Bong et al. only use three measurement settings and a reduced Bell inequality. 

In a very recent paper by Moreno et al., the authors tried to quantify $(AOE)$ and showed that `observed facts are maximally non-absolute' by invoking versions of generalised Bell inequalities.\cite{Moreno2022eventsinquantum} This is in the spirit of our claim, but the conclusion of our result is about the predictability of observers rather than the `absoluteness of facts': If reliable intersubjectivity---i.e. the more general version of $(AOE)$---doesn't hold, then Maggie can in principle have predictive advantage at the cost of it being private and hence `non-absolute'. This kind of improved predictivity therefore introduces an epistemic boundary between Maggie and other observers.\footnote{Note further that Moreno et al introduce a relaxation of $(AOE)$ which they call $\epsilon$-AOE and it is the assumption that for the special measurement that Cavalcanti consider (the record checking basis) the probability that the two outcomes are equal is not 1 but 1-epsilon. This again can be related to a special choice of predictive advantage where for the record checking basis the predictions are probabilistic. Thus, their assumption is related to predictive advantage but not as general as how it is defined here.}

\paragraph{Colbeck-Renner Theorem.} Another important result by Colbeck and Renner purportedly establishes that `no extension of quantum theory can have improved predictive power' (even for arbitrary quantum states and individual systems).\citep{colbeck2011no} If the claim holds up to scrutiny in the framework presented, it seems to be a robust result on the empirical completeness problem. 

The proof has been analysed and criticised, e.g. in \cite{HERMENS2020121, Leifer2014extended, LEEGWATER201618, Landsman, Ghirardi-on-Colbeck-Renner}. The criticism is mainly about Colbeck and Renner's assumptions implying parameter independence (which is much stronger than assuming no-signalling alone), the status of `free will' and its relationship to the central premise, and, moreover, lacking the foundation for addressing individual systems for which much stronger assumptions seem necessary.

The Colbeck-Renner theorem proceeds on two basic assumptions. In prose, `Measurement settings can be chosen freely: `[...] the input, A, of a measurement process can be chosen such that it is uncorrelated with certain other spacetime random variables, namely all those whose coordinates lie outside the future light cone of the coordinates of A.', in conjunction with the empirical validity of quantum predictions: `Measurement outcomes obey quantum statistics, and all processes within quantum theory can be considered as unitary evolutions if one takes into account the environment.'\footnote{They add that this only needs to hold for `microscopic processes on short timescales.'}

The main result is then stated as the quantum probabilities being `the most accurate prediction'  of measurement outcomes. That is, for any system measured with setting $a$ and outcomes $x$, no additional information exists improving the quantum probability prediction $p(x|a)$. At this point, Colbeck and Renner mention that such additional information provided by a hypothetical `extension' of the theory is \textit{accessible} at any time.

Unfortunately, Colbeck and Renner do not address unambiguously in what sense predictive advantage and `extensions' of quantum theory are to be understood. That is, they don't deal with the quantum decorrelation principle or point out in what regime if any, quantum predictions have to be violated.\footnote{On the face of it, based on the two assumptions, their claim on empirical completeness appears even paradoxical. Recall the quantum decorrelation principle saying that a target system decouples from its environment whenever a pure state is assigned. If `no extension of quantum theory can have improved predictive power' \textit{while} preserving the standard quantum predictions, it is unclear what such an `extension' is supposed to look like.}

Furthermore, Colbeck and Renner do not address the epistemic nature of the variables (here Maggie's records $z$), and, importantly, how they relate to the variables of other observers---as is established here by explicitly assuming $(RI)$. This confuses their analysis, for it is hence not entirely clear if the result pertains to controllable or uncontrollable variables. The claims crucially hinge on whether there exists an epistemic boundary between Maggie and Alice.\footnote{Consequently, it depends on the details of what is assumed about the variables' relationship to one another whether the locality assumption amounts to no-signalling or parameter-independence, and whether the variables are even \textit{uniquely defined} for all observers. In the accounts of Hermens and Leifer, the theorem is restated in the context of ontological models, providing an unambiguous underpinning of the result. There the `free-choice' assumption (FR) straightforwardly translates to parameter independence (and settings independence). This is implied by the fact that the relevant variables enter the story as the \textit{full} ontic (non-manifest) state $z=\lambda$ of a preparation determining all outcome probabilities of arbitrary measurements. This seems to be the view as standardly conceived (see, for instance,  \cite{Leifer2014extended,Ghirardi-on-Colbeck-Renner,HERMENS2020121,LEEGWATER201618,Landsman}). Then again, the problem remains whether $z$ has anything to do with an actual measurement record that can be communicated to other observers. }

For example, the authors purport that their assumptions rule out pilot wave theory---which is true since (FR) does imply parameter independence (and is violated in Bohmian theories). But it is worth mentioning that pilot wave theory is no-signalling while still being parameter \textit{dependent}. Hidden variable theories are not a priori ruled out by the assumptions made in our approach. Regarding pilot wave theory our claim shows that the Bohmian configurations (if assumed to exist) cannot be jointly known by Maggie and Alice (although $z$ may still represent a Bohmian particle configuration) if they are to deterministically fix all measurement outcomes.\footnote{See also the idea of a coarse-grained hidden variable in Ghirardi and Romano's approach, which resembles the thought that the $z$ variables don't necessarily refer to the full ontic states \citep{Ghirardi-Romano-inequivalent}.} 

\paragraph{Violations of Reliable Intersubjectivity.} Intriguingly, Maggie can in principle know the outcomes of observables corresponding to non-commuting operators, but wouldn't be able to communicate that to Alice. As long as reliable intersubjectivity is violated, she can predict deterministically the outcomes $x$ of Alice's measurement for arbitrary settings $a$ (see pilot wave theory below). Nevertheless, the result shows that Maggie cannot communicate that knowledge to Alice since that would invariably disturb Maggie's record of $z$. In Everettian quantum mechanics, a similar story can be told (see Everettian Quantum Mechanics below). However, there is no disturbance of the memory involved since the possible outcomes exist already in different branches, and Maggie's wave function branch will match the corresponding branch of Alice. In this case, reliable intersubjectivity is violated by introducing non-unique outcomes, i.e. $z$ doesn't possess a single definite value.

\paragraph{Pilot Wave Theory.}

In (equilibrium) pilot wave theory \cite{Bohm-pilot-wave, goldstein-sep-qm-bohm} the particle configurations are non-quantum variables in the sense that in general they do not satisfy the quantum decorrelation principle---e.g. the particle having a definite position doesn't imply that the corresponding wavefunction has very narrowly peaked support.

As it turns out, it is no contradiction for an observer to obtain Bohmian particle configurations, even in the cases where they behave non-classically. That is, in a Bohmian world it is in principle possible for Maggie to possess a magic prediction box that spits out the Bohmian initial configurations of the setup, therefore having deterministic predictive advantage. But upon no-signalling---which is satisfied in pilot wave theory on the level of manipulable variables---her records cannot be reliably reported to Alice. This follows from the observation that A) Maggie can in principle measure the Bohmian configurations represented by $z$, and they would be perfectly correlated to Alice's measurement outcomes, but B) if Maggie's records could be reliably communicated to Alice, then Alice could use them to signal superluminally to Bob (since Bohmian configurations are non-local). Thus, although Maggie may be able to observe them initially and make an improved prediction, her records must be disturbed through interaction with Alice to recover the correct quantum statistics on the level of Alice. This should hold for any kind of super-quantum predictions. In the context of hidden variable theories the interaction between Maggie and Alice thus always introduces disturbance whenever the variable $z$ represents \textit{any} degree of predictive advantage. Reliable intersubjectivity must thus be false in such theories to restore no-signalling on the operational level.  Brukner et al. came to a similar conclusion for a special measurement scenario involving two measurements, and showed a no-go result for `the persistent reality of Wigner's friend's memory'.\cite{Brukner-persistent-reality-of-Wigner} 

Whenever Maggie's variable would constitute some pre-determination of Alice's outcomes beyond the quantum probabilities, its value will be concealed from Alice. They cannot both live in a mutual `reality' where the record $z$ exists for both simultaneously. This epistemic boundary between observers is the mechanism by which `hidden variables' are hidden. 


\paragraph{Everettian Quantum Mechanics.} In Everettian quantum mechanics \cite{Everett-book, Everett1973}, reliable intersubjectivity is violated due to measurement records not being unique.  Again the question arises if this could constitute predictive advantage.  The variables that are in principle available to an observer like Maggie are the total wave function of the setup and knowledge of the branch she is in. Conditioned on the total wave function alone Maggie's means to predict Alice's outcomes are limited by the Born probabilities. Thus, the question is whether knowledge of the subjective branch could improve her predictions. In certain scenarios Maggie can indeed have \textit{relative} predictive advantage. For instance, Maggie may have measured a quantum system with two outcomes where after the interaction there exist two copies of Maggie having recorded one of the outcomes. The total wavefunction is still a superposition of those two branches, but if Alice measures in the same basis---i.e. checking Maggie's record---each Maggie can predict with certainty that there will be an Alice observing the same outcome, and, thus each Maggie has relative predictive advantage. It would be interesting to see what the conclusions are for other choices of Alice's basis. We leave that for further investigations. 

We suspect that analogous conclusions can be drawn in Rovelli's `relational quantum mechanics' \cite{Rovelli-RQM} and pragmatist views of quantum mechanics like QBism \cite{sep-quantum-bayesian}.

\paragraph{Laplace's Demon.}The epistemic boundaries between observers resulting from a failure of $(RI)$ afford an analogy in the spirit of Laplace’s demon in classical physics. The demon has to be an isolated creature external to the universe that it is describing in order to avoid logical inconsistencies whenever it is itself part of the system. Although there exists no direct link to a self-referential problem in the quantum case, the situation is similar in that quantum observers could exist with predictive advantage (even deterministic) but who are epistemically separated from the system and other observers. Moreover, when the different observers interact, to keep the epistemic boundary intact, reliable intersubjectivity is violated such that either observed records can have well-defined values and Maggie's variable $z$ (and her memory states) has to flip, or, the observed records of $z$ are subjective and not well-defined values, but the variable $z$ is preserved when communicated to Alice.

\section{Conclusions}
\label{sec:conclusion}
Can post-quantum theories have predictive advantage whilst conforming to the Born rule on average? We showed that unless epistemic boundaries exist between observers, the answer is no: quantum uncertainty is an unavoidable limit to Nature’s predictability. For any hypothetical predictive advantage, there exists a quantum measurement scenario with a finite number of settings which implies that predictive advantage would violate no-signalling. Giving up the assumption of `reliable intersubjectivity' does in principle enable (subjective) predictive advantage (and even knowledge about the outcomes of non-commuting observables), but implies an epistemic boundary between different quantum observers. 

To our knowledge, this is the first rigorous result on predictive completeness addressing locality and the mutual epistemic reality of macroscopic observers. 


We left open the question whether (operational) principles---other than no-signalling---can be invoked for a claim about empirical completeness. For example, these could be dynamical or methodological assumptions about the prediction improving variables. Moreover, no general proof is currently available for arbitrary single systems. The impossibility of predictive advantage for single systems may be established on grounds of non-contextuality---as a generalisation of locality. 

In sum, if we assume that observers can reliably communicate and can't send superluminal signals, the contemporary mainstream view that quantum predictions are irreducibly probabilistic becomes a precise theorem. Consequently, not only can quantum mechanical measurement outcomes not be predicted with certainty, the no-signalling predictions of \textit{any} theory must be exactly the quantum probabilities.

Insofar as we hold on to the principles of relativity and intersubjectivity, it seems fair to say that one more thing we may add to the very few things we can be certain about: \textit{the certainty of uncertainty.}

\section*{Acknowledgments}

We thank Christopher Timpson, Owen Maroney, Guido Bacciagaluppi, Jonathan Barrett, Tom\'{a}\v{s} Gonda, and Y\`il\`e Yīng for helpful comments. This research was funded in part by the Austrian Science Fund (FWF) 10.55776/Y1261. For open access purposes, the author have applied a CC BY public copyright license to any author accepted manuscript version arising from this submission.

\bibliographystyle{quantum}
\bibliography{library}

\appendix
\section{Methods}
\label{app:methods}

The proof of Claim \ref{eqn:main result} builds on so-called chained Bell inequalities \cite{BRAUNSTEIN-caves-chained-bell-inequalities}, and uses a result established by Barrett, Kent, and Pironio \cite{Barrt-Kent-Pironio-local-decomposition}.  

We start by noting that the system Alice and Bob share is assumed to be a bi-partite quantum system represented by a density operator $\rho\in\mathcal{S}(\mathcal{H}\otimes\mathcal{H})$ as the (mixed) state of the system, where $\mathcal{S}(\mathcal{H}\otimes\mathcal{H})$ is a convex subset of the self-adjoint trace-class operators on $\mathcal{H}=\mathbb{C}^2\otimes\mathbb{C}^2$.  We will only consider pure states of the form 
\begin{equation}
	\label{eqn:arbitrary entangled bipartite state}
	\ket{\Psi}=\alpha\ket{00'}+\sqrt{1-\alpha^2}\ket{11'}, 
\end{equation} and corresponding density operator $\rho=\ket{\Psi}\bra{\Psi}$. A subsystem's state is then recovered by the partial trace $\rho_S=\tr_{S'}(\rho)$. 

We consider separable POVM measurements $E_x^a, E_y^b$ for both Alice and Bob with quantum probabilities
\begin{equation}
	p(x,y|a,b,\rho):= \tr(E_x^a \otimes E_y^b\rho).
\end{equation}  Every such POVM element can be written as 
\begin{equation}
	E_i= \frac{1}{2}(\alpha_0^i\mathds{1}+\bm{\alpha}^i\cdot\bm{\sigma}),
\end{equation} where $\alpha_0^i$ a real number,  $\bm{\alpha}^i=(\alpha_x^i,\alpha_y^i,\alpha_z^i)$ a normalised vector, and $\bm{\sigma}=(\sigma_x, \sigma_y, \sigma_z)$ the Pauli matrices spanning the space of self-adjoint operators on $\mathbb{C}^2$. 

It will suffice to consider projective measurements in the plane orthogonal to the direction in which the two qubit systems travel. Formally, this means we can parametrise $E_i=\frac{1}{2}(\mathds{1}+(-1)^i\sigma_n)$, where 
\begin{equation}
	\sigma_n:= \hat{n}\cdot\vec{\sigma}= 
	\begin{pmatrix}
		\cos\vartheta & \sin\vartheta  \\
		\ \sin\vartheta  & -\cos\vartheta \\
	\end{pmatrix}, 
\end{equation} for the measurement axis $\hat{n}=(\sin\vartheta, 0, \cos\vartheta)$. We shall denote Alice's setting with $\vartheta_1$ and Bob's with $\vartheta_2$, respectively. Moreover, the two observers are supposed to be able to each perform measurements in $N$ different settings. 

Suppose Maggie's prediction box consists of a pointer yielding values for the variable $z$ for each run of the experiment. Suppose further that $z$ is distributed according to the measure $\mu(z|a,b,\rho)$, which generally depends on the measurement settings $a,b$ and the prepared quantum state $\rho$. 

$(RI)$ allows to define an objective probability distribution $p(x,y|a,b,\rho, z)$ which is well-defined for all observers in the setup. Furthermore, (SD) ensures that all predictions of the three agents supervene on accessible variables and that all variables used are operationally manipulable. The quantum probabilities are reproduced if 
\begin{align}
	p(x,y|a,b,\rho)&=\tr(E_x^a\otimes E_y^b \rho)\\ \nonumber
	&= \int p(x,y|a,b,\rho, z)\mu(z|\rho)dz,
\end{align} which we assume in the following. We have also made use of $(NC)$ which implies that $z$ is independent of Alice's and Bob's measurement settings. Without reliable intersubjectivity it wouldn't be possible to ascribe a unique value to the settings $a, b$, and, in particular, the outcomes $x, y$ and Maggie's variable $z$.  The values might be perspectival, i.e. depend on some other parameter in the experimental setup like the observers. It could also be that the measured outcomes don't take on single values. As explained earlier, $(RI)$ is violated notably in many worlds and relational interpretations. Moreover, reliability was also argued to be violated in some classes of hidden variable theories. There is no guarantee that Maggie's outcomes are objectively defined for all observers, but importantly, there is also no guarantee that if they are, they are accessible to observers other than Maggie. For example, in pilot wave theory the configurations describing Maggie depend on the total wavefunction of the setup, and thus on Alice's measurement by virtue of the dynamical laws for the particle configurations. Interestingly, the theorem shows that for predictive advantage of \textit{any kind} and all measurements settings of Alice a contradiction arises. 

Note that this assumption is necessary for the proof, but is not explicit in other no-go theorems such as Brukner et al.'s and Cavalcanti et al.'s results.\cite{Brukner-Wiger-no-go, Cavalcanti-local-friendliness-no-go} (see also the same ambiguity in the definition of `cross-perspective links' in \cite{Adlam-Rovelly-information-is-physical}.)

Furthermore, upon $(RI)$ the marginals are well defined, i.e.
\begin{equation}
	p(x|a,b,\rho, z) = \sum_y p(x,y|a,b,\rho,z).
\end{equation} These are the predictions of Maggie for hypothetical predictive advantage over Alice's outcomes. We assume that Maggie only needs to make predictions about Alice's outcomes, not for the probabilities of the total system consisting of Alice plus Bob. 

For the special case of Bell's experiment suppose the predictive advantage consists of deterministic predictions for all measurements. Let $p(x,y|a,b,z)$ be the predictions of an empirical extension. A deterministic prediction entails that the outcomes $x,y$ are functions of the settings $a,b$ and extension variable $z$. That is, $(x,y)=(f(a,b,z), g(a,b,z))$, and so 
\begin{align}
	\label{eqn:deterministic predictions}
	p(x,y|a,b,z)&=\delta_{(x,y),((f(a,b,z),g(a,b,z)))}\\ \nonumber
	&=\delta_{(x),f(a,b,z)}\delta_{(y),g(a,b,z)}\\ \nonumber
	&=p(x|a,b,z)p(y|a,b,z). 
\end{align} But when on top of that, the probabilities are to satisfy no-signalling, then 
\begin{align}
	\label{eqn:signal-local deterministic predicitons}
	p(x,y|a,b,z)&=p(x|a,b,z)p(y|a,b,z)\\ \nonumber
	&=p(x|a,z)p(y|b,z). 
\end{align} Thus, the predictions describe uncorrelated and no-signalling correlations that neither depend on the outcomes of the other site nor its settings. However, such cannot reproduce all quantum correlations due to Bell's theorem. Therefore, $p(x,y|a,b,z)$ cannot be (fully) deterministic (cf. also \citep{PR-boxes}, \citep{Masanes}, \citep{Cavalcanti-predictability}, and \citep{Fankhauser-thesis}). Thus, we conclude that any empirical extension of quantum mechanics necessarily makes indeterministic predictions for at least some settings and outcomes. This presents a first step for the result of this work, but note that Bell's theorem does \textit{not} provide proof the empirical completeness of quantum theory. What we demanded from a theory with predictive advantage was the possibility of finding \textit{some} decomposition of quantum probabilities. That is, it need not be deterministic in general, and, in particular, not necessarily deterministic for the outcomes of \textit{all} measurements. In fact, for a Bell setup with two measurement settings on each site probabilistic theories with predictive advantage can be constructed.\cite{Ghirardi-Romano-inequivalent}

In the next step we study the probability decompositions with generalised Bell measures, a version of which reads
\begin{align}
	I_N:= &\langle|x_{\vartheta_1}-y_{\vartheta_1}|\rangle+\langle|y_{\vartheta_1}-x_{\vartheta_2}|\rangle
	+ \langle|x_{\vartheta_2}-y_{\vartheta_2}|\rangle\\
	&+ ... + \langle|x_{\vartheta_N}-y_{\vartheta_N}|\rangle+\langle|y_{\vartheta_N}-x_{\vartheta_1}-1|\rangle,\nonumber
\end{align} where the expectation value is defined as $\langle A\rangle:=\sum_{i=0}^{1} i p(A=i)=p(A=1)$. By defining $x_{\vartheta_{N+1}}:=x_{\vartheta_1}+1 ~(\bmod~ 2)$ it can be written as $I_N=\sum\limits_{n=1}^N  (\langle|x_{\vartheta_n}-y_{\vartheta_n}|\rangle+\langle|y_{\vartheta_n}-x_{\vartheta_{n+1}}|\rangle)$. Up to minor details this expression corresponds to the so-called chained Bell inequalities first introduced by \citep{BRAUNSTEIN-caves-chained-bell-inequalities}. See in particular the generalisation to arbitrary dimensions in \citep{Barrt-Kent-Pironio-local-decomposition} who we closely follow in the subsequent presentation. This is the generalised Bell inequality.\footnote{For $N=2$, $I_N\geq 1$ coincides with the familiar CHSH inequality.}  

We can now compute the value of $I_N$ for the states and settings introduced. For the expectation values in the sum, only those quantities are relevant for which $|x_{\vartheta_k}-y_{\vartheta_l}|=1$. Since all outcomes can be $0$ or $1$, this is satisfied whenever the outcomes are unequal. There are two cases for which this is the case, and the corresponding measurements are $E_1^{\vartheta_k}\otimes E_0^{\vartheta_l}$ and $E_0^{\vartheta_k}\otimes E_1^{\vartheta_l}$. Each term (except the last) in the sum is therefore given by
\begin{align}
	\langle|x_{\vartheta_k}-y_{\vartheta_l}|\rangle&=p(|x_{\vartheta_k}-y_{\vartheta_l}|=1)\\ \nonumber
	&=\tr(E_1^{\vartheta_k}\otimes E_0^{\vartheta_l}\rho)+\tr(E_0^{\vartheta_k}\otimes E_1^{\vartheta_l}\rho). 
\end{align} This holds for all terms but the last, where the argument is $1$ if and only if both outcomes are equal. For the latter case, the relevant probability is thus
\begin{dmath}
	\langle|x_{\vartheta_k}-y_{\vartheta_l}-1|\rangle=p(|x_{\vartheta_k}-y_{\vartheta_l}-1|=1)\\ \nonumber
	=\tr(E_0^{\vartheta_k}\otimes E_0^{\vartheta_l}\rho)+\tr(E_1^{\vartheta_k}\otimes E_1^{\vartheta_l}\rho). 
\end{dmath} 

Summing up gives the generalised Bell measure for the entangled state \ref{eqn:arbitrary entangled bipartite state} with $N$ arbitrary measurement settings for Alice and Bob, i.e.
\begin{dmath}
	\label{eqn:general Bell measure for arbitrary states}
	I_N=\sum\limits_{n=1}^N  (\langle|x_{\vartheta_n}-y_{\vartheta_n}|\rangle+\langle|y_{\vartheta_n}-x_{\vartheta_{n+1}}|\rangle)\\
	=\sum_{n=1}^{N}\left[\sin^2(\frac{\vartheta_n}{2}-\frac{\vartheta_n'}{2})+\sin^2(\frac{\vartheta'_n}{2}-\frac{\vartheta_{n+1}}{2})\right]\\ \nonumber
	-(\alpha\sqrt{1-\alpha^2}-\frac{1}{2})\sum_{n=1}^{N}\left[\sin\vartheta_n\sin\vartheta_n'+\sin\vartheta_n'\sin\vartheta_{n+1}\right].\nonumber
\end{dmath}

Interestingly, $I_N$ can be related to the marginal probabilities for Alice's outcomes, and, as a result, to the hypothetical predictions of Maggie. This crucial insight was observed by \citep{Barrt-Kent-Pironio-local-decomposition}. Barrett et al.'s result requires two assumptions. First, the probabilities are no-signalling---which was introduced as no-signalling $(SL)$ here)---, and setting independence---the assumption of no-conspiracy $(NC)$. Thus, we can apply their result to the probability assignment of Alice's outcomes conditioned on Maggie's variable.  

With these assumptions we arrive at the claim that the difference between the two probabilities for Alice's outcomes is bounded from above by the $I_N$, i.e.
\begin{equation}
	I_N\geq |p(x|a=\vartheta_k,z)-p(1-x|a={\vartheta_k},z)|
\end{equation} for any choice of setting $\vartheta_{k}$  (cf. \citep{Barrt-Kent-Pironio-local-decomposition}, and also \citep{Fankhauser-thesis}). 

In the quantum case, the difference would vanish for maximally entangled states if Maggie didn't have any predictive advantage. For Alice's outcomes would be equally likely. Thus, on the assumption of predictive advantage $(PA)$ it follows that the difference cannot vanish. That is, for all predictions of Maggie there exists an $\varepsilon\in\mathbb{R}$ such that
\begin{equation}
	\label{eqn:predictive advantage epsilon}
	|p(x|a=\vartheta_k,z)-p(1-x|a={\vartheta_k},z)|> \varepsilon.
\end{equation}

Maggie's decomposition of probabilities has to recover the quantum probabilities on average, and so one must have
\begin{align}
	&\int|p(x|a=\vartheta_k,z)-p(1-x|a={\vartheta_k},z)|\mu(z)dz\\ \nonumber
	&\leq \int I_N \mu(z)dz= I_N(QM). 
\end{align} Thus, upon assumptions (1)-(5) the predictive advantage of Maggie is bounded from above by the Bell measure $I_N(QM)$ for the quantum correlations.  

We can look for measurement scenarios where the quantum value is minimised to find limits of Maggie's predictive advantage, i.e.
\begin{equation}
	\min\limits_{N, \vartheta_1, ..., \vartheta_N,\vartheta'_1, ... ,\vartheta_N} I_N.
\end{equation} 

Equation \ref{eqn:general Bell measure for arbitrary states} contains a term that only depends on the difference of measurement angles and is independent of the parameter $\alpha$. The second term vanishes when the state is maximally entangled, i.e. $\alpha=\frac{1}{\sqrt{2}}$. In this case, one can then indeed find a set of measurements for which the generalised Bell measure contradicts the existence of predictive advantage, i.e. Equation \ref{eqn:predictive advantage epsilon}.

We can choose a set of measurement angles for which the interval $\left[0,\pi\right]$ is divided into $N:=\left \lceil{\frac{\pi^2}{4\varepsilon}}\right \rceil $ equally spaced settings for both Alice's and Bob's measurement. That is, define, for example, the primary setting as 
\begin{equation}
	\vartheta_k:=\frac{\pi}{N}(k-1), k=1,..., N,
\end{equation} and the secondary setting as 
\begin{equation}
	\vartheta_l:=\frac{\pi}{N}(l-\frac{1}{2}), l=1,..., N.
\end{equation} With these definitions each term in $I_N$ is identical and reads $\sin^2\frac{\pi}{4N}$. Hence, there are $2N$ equal terms in the sum and 
\begin{equation}
	I_N\leq 2N\left(\frac{\pi}{4N}\right)^2\leq\frac{\varepsilon}{2}< \varepsilon
\end{equation}  (cf. also \citep{Barrt-Kent-Pironio-local-decomposition, ELITZUR199225}). Note that this statement is robust against some upper error bound regarding the accuracy to which quantum probabilities are confirmed in the experiment. That is, without having to reproduce quantum statistical data exactly, an actual measurement of $I_N$ still restricts the amount of predictive advantage up to its measurement accuracy. 

We conclude that the assumptions (1)--(5) can thus not all be jointly true, and that local observers are bound to the quantum uncertainty in terms of making predictions about outcomes of measurements.

\end{document}